\documentclass[a4paper]{article}
\usepackage{INTERSPEECH2021}

\usepackage{amsmath,graphicx}
\usepackage{acronym}
\usepackage{cite}
\usepackage{subcaption, tabularx}
\usepackage{url}
\usepackage{booktabs,balance}
\usepackage{pifont}
\usepackage{xcolor}

\newcommand{\cmark}{\ding{51}}%
\newcommand{\xmark}{\ding{55}}%

\acrodef{STFT}{short-time Fourier transform}
\acrodef{ASR}{automated speech recognition}
\acrodef{NN}{neural network}
\acrodef{MSE}{mean square error}
\acrodef{DFT}{discrete Fourier transform}
\acrodef{BN}{batch-normalisation}
\acrodef{PESQ}{perceptual evaluation of speech quality}
\acrodef{WPE}{weighted linear prediction error}
\acrodef{TDOA}{time difference of arrival}
\acrodef{RIR}{room impulse response}
\acrodef{RMS}{root-mean-square}
\acrodef{DSS}{deep symmetric sets}
\acrodef{GAN}{generative adversarial network}
\acrodef{SNR}{signal-to-noise ratio}
\acrodef{DOA}{direction of arrival}
\acrodef{CD}{cepstral distance}
\acrodef{PESQ}{perceptual evaluation of speech quality}
\acrodef{FWSegSNR}{frequency weighted segmental SNR}

\title{Scene-Agnostic Multi-Microphone Speech Dereverberation}
\name{Yochai Yemini$^1$, Ethan Fetaya$^1$,  Haggai Maron$^2$, Sharon Gannot$^1$}
\address{
  $^1$Faculty of Engineering, Bar-Ilan University, Ramat-Gan, 5290002, Israel\\
  $^2$NVIDIA Research}

\email{
\{Yochai.Yemini,Ethan.Fetaya,Sharon.Gannot\}@biu.ac.il,hmaron@nvidia.com
}

\begin{document}
%
\maketitle 
\begin{abstract}
\Acp{NN} have been widely applied in speech processing tasks, and, in particular, those employing microphone arrays. Nevertheless, most existing NN architectures can only deal with fixed and position-specific microphone arrays. In this paper, we present an NN architecture that can cope with microphone arrays whose number and positions of the microphones are unknown, and demonstrate its applicability in the speech dereverberation task. To this end, our approach harnesses recent advances in deep learning on set-structured data to design an architecture that enhances the reverberant log-spectrum. We use noisy and noiseless versions of a simulated reverberant dataset to test the proposed architecture. Our experiments on the noisy data show that the proposed scene-agnostic setup outperforms a powerful scene-aware framework, sometimes even with fewer microphones. With the noiseless dataset we show that, in most cases, our method outperforms the position-aware network as well as the state-of-the-art \ac{WPE} algorithm. 

\let\thefootnote\relax\footnotetext{This project has received funding from the European Union's Horizon 2020 Research and Innovation Programme, Grant Agreement No.~871245.
}

\end{abstract}
\section{Introduction}
As a sound wave propagates in an acoustic enclosure, it reflects from the room's facets and from the objects within the room. A microphone in that room will capture both the direct path signal and the associated reflections. This phenomenon, known as reverberation, negatively impacts speech quality and, in severe cases, even its intelligibility \cite{intro1}, posing difficulties to hearing impaired people as well as to \ac{ASR} systems. With the soaring popularity of \ac{ASR} based agents such as Amazon Alexa, Microsoft Cortana, Google Assistant and Apple Siri, mitigating reverberation is further incentivised.

In many cases, we have access to several audio signals that are captured by different microphones. For example, any modern mobile-phone is equipped with at least two microphones. Importantly, having access to several recordings of the same audio signal enables algorithms to reinforce spectral cues with spatial information, leading to better results compared to the single microphone case. For approaches that require a training stage, there may exist two main inconsistencies between the training and test conditions with respect to: (1) the number of microphones (2) the geometry and the position of the microphone array.

In realistic scenes, the number of microphones can vary over time. Usually, learning-based paradigms cannot straightforwardly accommodate such modifications to the scene. In order to be invariant to the number of microphones, several systems have to be trained, each for a specific number of microphones.
The relative positions of the microphones may also change. In a \emph{position-aware} setup they are constant and known, e.g. a fixed microphone array. In a \emph{position-agnostic} setup we do not have any prior information on the relative positions of the microphones. For example, this is the case when several different people capture the same audio signal spontaneously at a concert or a lecture \cite{smaragdis}. We dub an algorithm which is both position-agnostic and invariant to the number of microphones as \emph{scene-agnostic}.

In general, traditional speech processing algorithms are designed to deal with both scene-aware and scene-agnostic cases. In the context of speech dereverberation, a plethora of methods exists \cite{Naylor, REVERB}. Several approaches use an estimation scheme for filter coefficients. In the \ac{STFT} domain, reverberation can be modelled as a convolution along frames between the clean STFT and a reverberation filter, independently for each frequency.  In \cite{WPE}, the inverse of this filter is estimated by minimising the \ac{WPE}. In \cite{boaz}, the authors follow a iterative expectation-maximization paradigm to directly estimate the reverberation filters and to jointly dereverberate the signal using the Kalman filter algorithm. A different approach \cite{cauchi, boaz2} deploys a late reverberation analysis to devise a dereverberation system. 

In this paper  we tackle the more general scene-agnostic setup for \ac{NN}-based algorithms. We wish to harness the power of \acp{NN}, while circumventing their drawbacks. In the proposed scheme, the network takes as an input several distorted audio signals represented as log-spectra and predicts a single high quality spectrogram of the underlying audio signal. The main challenge here is devising an \ac{NN} architecture that is suitable for the scenario at hand. There are two main requirements: first, the architecture should be \emph{invariant} to the number and order permutations of the signals; second, it should  respect the fine details of the spectrogram image, namely to be able to reconstruct its complex structures such as the pitch and formants.

Previous approaches successfully dealt with only one of these requirements: for example, the multi-microphone setup is often dealt with by concatenating the per microphone \ac{STFT} along the channel axis \cite{wang_mc_dereverb, mc1, mc2}. This approach is useful when the array geometry is fixed. However, permuting the indices within the array while maintaining the positions or changing the respective geometry of the microphones and their positions, will lead to a mismatch with respect to the training data, and possible degradation in reconstruction quality. Another critical setback that this architecture raises is its incompatibility to a variable number of microphones. A recent line of studies has taken steps to mitigate this problem by adopting the \emph{Deep Sets} \cite{zaheer} paradigm for speech separation, e.g. \cite{TAC, wang2020neural}. 



In this work, dereverberation is carried out via log-spectrum enhancement while the phase remains unaltered, rendering the problem an image-like estimation. To address both challenges mentioned above, we leverage a recently suggested architecture for learning sets of symmetric elements \cite{sets}. This framework captures both the set symmetry and the intrinsic symmetries of each element in the set by replacing  convolution layers with \ac{DSS} layers. 

To demonstrate the efficacy of our approach we provide an extensive experimental evaluation on a new simulated dataset. We strive for making this dataset publicly available, to encourage further research in the field of scene-agnostic NNs. 
The new dataset comprises several realistic scenarios that represent different positioning of the microphones in various rooms and reverberation conditions. 

\section{Problem Formulation}
\label{sec:prob_form}
Let $x(t)$ denote the anechoic signal in the discrete-time domain. The reverberant signal can be described as the convolution between $x(t)$ and the reverberant \ac{RIR}. Therefore, given a microphone array with $M$ microphones, the received signal at each microphone is 
\begin{equation}
    y_i(t) = \{x\ast h_i\}(t) + n_i(t), \quad i=1,2,\ldots,M
\end{equation}
where $h_i(t)$ and $n_i(t)$ are the per microphone \ac{RIR} and low-level stationary noise, respectively. The objective is to estimate $x(t)$ given the corrupted observations $\{y_i(t)\}_{i=1}^M$.

To achieve this goal, the signals are first transformed to the log-spectrum representation. Let $Y_i(n,k)$ denote the log-spectrum of $y_i(t)$, i.e.~the log-absolute value of the \ac{STFT} of $y_i(t)$ at the $n$-th frame and the $k$-th frequency bin, where $k=0,1,\ldots,K-1$. Due to the symmetry of the \ac{DFT}, only the first $K/2+1$ frequency elements are considered. Denote  $F=K/2$. For brevity, the time and frequency indices will be omitted when they are unnecessary for the clarity of the paper.

In this work, an \ac{NN} is used to predict $X$, the log-spectrum of the clean speech signal. After the enhanced log-spectrum is obtained, it is transformed back to the time domain using the noisy phase of the \ac{STFT} from the microphone that recorded the signal with the largest power. We have two requirements from the network. Firstly, for a specific value of $M$, the network's output must remain the same for any permutation of $\{Y_i\}_{i=1}^M$. Secondly, we want a single network to be capable of processing $\{Y_i\}_{i=1}^M$ for several values of $M$. 

Note that na\"{i}ve concatenation of the multi-microphone input along the feature axis of the network does not guarantee any sort of invariance, neither with respect to the number of microphones nor their positions. Therefore, we expect it to be less suitable for scene-agnostic processing.
\begin{figure}[t!]
    \centering
    \includegraphics[scale=0.4]{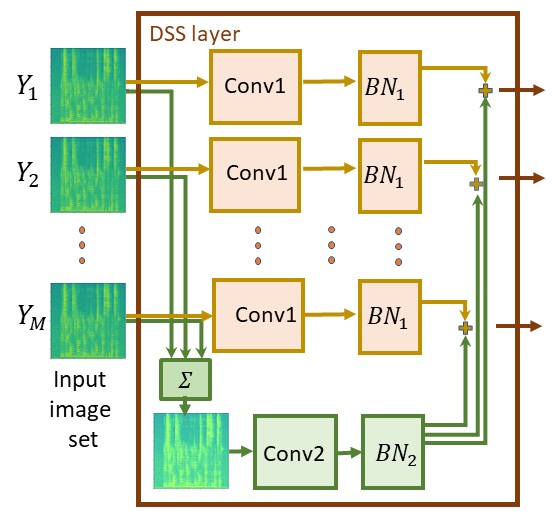}
    \caption{The \ac{DSS} layer \cite{sets} used in this work, composed of a Siamese network applied independently to each image and an aggregation branch, realised as a sum operation. The Siamese network and the aggregation branch use distinct convolution and \ac{BN} layers. }
    \label{fig:dss}
\end{figure}
\section{Algorithm}
The proposed method is inspired by several studies. It is based on dereverberation via image deblurring in the log-spectrum domain \cite{ori} using a U-net architecture \cite{unet} that estimates a clean spectrogram directly from the reverberated one. This approach demonstrated promising results, surpassing several baselines by a significant margin. We extend it to the multi-microphone case by deploying the powerful burst image deblurring framework \cite{Aittala}, which takes in a set of blurry images and outputs a clean image. A U-net architecture with \ac{DSS} layers \cite{sets} serves as our network of choice. The invariance property of the \ac{DSS} layer allows the network to process ad hoc distributed microphone arrays, in addition to fixed geometry arrays. The \ac{DSS} layers can be coupled with other architectures, such as the one in \cite{unet++}.

\noindent \textbf{Preprocessing: }
For a given set of reverberant time-domain samples $\{y_i(t)\}_{i=1}^M$, a preprocessing step first normalises the \ac{RMS} across all $M$ signals to a constant value of 0.1. Then, the log-spectra images $\{Y_i\}_{i=1}^M$ are computed. The $M$ log-spectra are divided to slices of 256 frames each, resulting in multiple slices of size $M\times 1\times 256\times (F+1)$, where the second dimension signifies the feature axis. Since the highest frequency alone does not carry much information, and for computational reasons, it is not enhanced. The $M$ images of size $1\times 256\times F$ are linearly mapped to the range $[-1, 1]$, which makes learning faster and more stable. 

\noindent \textbf{Network Architecture: }
Our network receives the $M\times 1\times 256\times F$ input, and predicts an output of size $1\times 256\times F$. After all slices have been enhanced, the estimated log-spectrum $\hat{X}$ is constructed by concatenating the enhanced slices along the time axis and then mapping the range $[-1, 1]$ back to the valid range of a spectrogram. Eventually, the highest frequency is reattached.
As mentioned, the proposed network is based on a U-net network and \ac{DSS} layers. Each \ac{DSS} layer receives and outputs a set of $M$ images (Fig. \ref{fig:dss}). Similarly to \cite{ori}, the U-net pools the spatial resolution of the input images to $1\times 1$, and then unrolls it back to the original resolution of $256\times F$. 

Most notably, this architecture can ingest different values of $M$ during training/test phases. This allows a single network to be applicable to arrays with a varying number of microphones. For this setup, the sum operation of the aggregation branch of the \ac{DSS} layer is replaced by mean.

\noindent \textbf{Implementation Details: }
The following notations are used to describe the network. A $\textrm{D}_{d/u,n}$ layer denotes a \ac{DSS} layer that comprises strided convolutions with a down/upsampling factor of 2 and has $n$ filters. L and R mark a LeakyReLU activation with a negative slope of 0.2 and a ReLU activation, respectively. All convolution layers use $4\times 4$ kernels. The encoder's layers are: \\
$\textrm{D}_{d,64} \textrm{L} \rightarrow \textrm{D}_{d,128} \textrm{L} \rightarrow
\textrm{D}_{d,256} \textrm{L} \rightarrow \textrm{D}_{d,512} \textrm{L} \rightarrow
\textrm{D}_{d,512} \textrm{L} \rightarrow \textrm{D}_{d,512} \textrm{L} \rightarrow
\textrm{D}_{d,512} \textrm{L} \rightarrow \textrm{D}_{d,512} \textrm{L}$ 
and the decoder's architecture is 
$\textrm{D}_{u,512} \textrm{R} \rightarrow \textrm{D}_{u,512} \textrm{R} \rightarrow
\textrm{D}_{u,512} \textrm{R} \rightarrow \textrm{D}_{u,512} \textrm{R} \rightarrow
\textrm{D}_{u,256} \textrm{R} \rightarrow \textrm{R}_{u,128} \textrm{R} \rightarrow
\textrm{D}_{u,64} \textrm{R} \rightarrow \textrm{D}_{u,1} \textrm{R}$. 
Skip connections are added between the encoder and decoder to form the U-net part of the network.
At the decoder's output, the dimensions are $M \times 1 \times 256 \times F$. In order to reach a set-based result, a max pooling layer is applied along the set dimension \cite{Aittala}, highlighting the dominant features. Eventually the network terminates with $\textrm{CBR}\rightarrow \textrm{CT}$, where $\textrm{C}$ is a convolution layer with stride 1 and a single output channel, $\textrm{B}$ stands for a \ac{BN} layer and T is the Tanh activation.

\begin{table*}[t]
    \centering
    \caption{Results for reverberant speech with a 20dB low-band noise for the different scenarios. The asterisk symbol represents a scene-agnostic architecture that was trained on both 4 and 8 microphones.}
    \label{table:BIUREV-N}
    \begin{subtable}[h!]{0.45\textwidth}
        \centering
        \setlength{\belowcaptionskip}{.1cm}
        \caption{Far}
        \label{table:far_noise}
        \resizebox{0.9\linewidth}{!}{%
        \begin{tabular}{@{}lcccc@{}}
        \toprule
         & Scene-Agnostic & \multicolumn{1}{c}{CD$\downarrow$} & \multicolumn{1}{c}{FWSegSNR$\uparrow$} & \multicolumn{1}{c}{PESQ$\uparrow$} \\
         \cmidrule(lr){3-3} \cmidrule(lr){4-4} \cmidrule(lr){5-5}
        {reverberant} & - & 5.92 & -0.74 & 1.20 \\
        \hline
        {1 mic. \cite{ori}} & \xmark & 3.37 & 8.64 & 1.40 \\ 
        {4 mics.} & \xmark & 3.06 & 9.98 & 1.61 \\
        {8 mics.} & \xmark & \textbf{2.94} & \textbf{10.35} & \textbf{1.74} \\
        \midrule
        {4 mics.} & \cmark & 3.12 & 9.86 & 1.64 \\
        {4 mics.*} & \cmark & 3.07 & 9.53 & 1.69 \\
        {8 mics.} & \cmark & 3.05 & 10.14 & 1.71 \\
        {8 mics.*} & \cmark & 3.01 & \textbf{10.35} & 1.71 \\
         \bottomrule
        \end{tabular}
    }%
    \end{subtable}%
    \begin{subtable}[h!]{0.45\textwidth}
        \centering
        \setlength{\belowcaptionskip}{.1cm}
        \caption{Near}
        \label{table:near_noise}
        \resizebox{0.9\linewidth}{!}{%
        \begin{tabular}{@{}lcccc@{}}
        \toprule
         & Scene-Agnostic & \multicolumn{1}{c}{CD$\downarrow$} & \multicolumn{1}{c}{FWSegSNR$\uparrow$} & \multicolumn{1}{c}{PESQ$\uparrow$} \\
         \cmidrule(lr){3-3} \cmidrule(lr){4-4} \cmidrule(lr){5-5}
        {reverberant} & - & 4.54 & 5.86 & 1.71 \\
        \hline
        {1 mic. \cite{ori}} & \xmark & 3.11 & 9.92 & 1.64 \\ 
        {4 mics.} & \xmark & 2.90 & 10.86 & 1.93 \\
        {8 mics.} & \xmark & 2.77 & 11.40 & 2.12 \\
        \midrule
        {4 mics.} & \cmark & 2.68 & 12.19 & 2.20 \\
        {4 mics.*} & \cmark & 2.55 & 12.19 & 2.35 \\
        {8 mics.} & \cmark & \textbf{2.52} & \textbf{12.80} & \textbf{2.39} \\
        {8 mics.*} & \cmark & 2.54 & 12.78 & 2.38 \\
        \bottomrule
        \end{tabular}}%
    \end{subtable}%
    
    \begin{subtable}[h!]{0.45\textwidth}
        \centering
        \setlength{\belowcaptionskip}{.1cm}
        \caption{Random}
        \label{table:random_noise}
        \resizebox{0.9\linewidth}{!}{%
        \begin{tabular}{@{}lcccc@{}}
        \toprule
         & Scene-Agnostic & \multicolumn{1}{c}{CD$\downarrow$} & \multicolumn{1}{c}{FWSegSNR$\uparrow$} & \multicolumn{1}{c}{PESQ$\uparrow$} \\
         \cmidrule(lr){3-3} \cmidrule(lr){4-4} \cmidrule(lr){5-5}
        {reverberant} & - & 4.95 & 2.96 & 1.52 \\
        \hline
        {1 mic. \cite{ori}} & \xmark & 3.25 & 9.19 & 1.53\\ 
        {4 mics.} & \xmark & 2.82 & 11.12 & 1.97 \\
        {8 mics.} & \xmark & 2.64 & 11.94 & 2.21 \\
        \midrule
        {4 mics.} & \cmark & 2.82 & 11.41 & 2.05 \\
        {4 mics.*} & \cmark & 2.77 & 11.07 & 2.09 \\
        {8 mics.} & \cmark & 2.62 & 12.30 & \textbf{2.30} \\
        {8 mics.*} & \cmark & \textbf{2.61} & \textbf{12.33} & 2.29 \\
        \bottomrule
        \end{tabular}}%
    \end{subtable}%
   \begin{subtable}[h!]{0.45\textwidth}
        \centering
        \setlength{\belowcaptionskip}{.1cm}
        \caption{Winning Ticket}
        \label{table:winning_noise}
        \resizebox{0.9\linewidth}{!}{%
        \begin{tabular}{@{}lcccc@{}}
        \toprule
         & Scene-Agnostic & \multicolumn{1}{c}{CD$\downarrow$} & \multicolumn{1}{c}{FWSegSNR$\uparrow$} & \multicolumn{1}{c}{PESQ$\uparrow$} \\
         \cmidrule(lr){3-3} \cmidrule(lr){4-4} \cmidrule(lr){5-5}
        {reverberant} & - & 4.82 & 3.5 & 1.5 \\
        \hline
        {\space\space\space\space-} & - & - & - & - \\
        {4 mics.} & \xmark & 2.70 & 11.73 & 2.13 \\
        {8 mics.} & \xmark & 2.66 & 11.89 & 2.22 \\
        \midrule
        {4 mics.} & \cmark & 2.64 & 12.31 & 2.27 \\
        {4 mics.*} & \cmark & 2.59 & 11.92 & 2.30 \\
        {8 mics.} & \cmark & \textbf{2.58} & \textbf{12.48} & \textbf{2.35} \\
        {8 mics.*} & \cmark & \textbf{2.58} & 12.43 & 2.34 \\
         \bottomrule
        \end{tabular}}%
   \end{subtable}%
\end{table*}
Unlike \cite{ori}, we only use square filters and do not incorporate any \ac{GAN} framework such as Pix2Pix \cite{pix2pix}. Based on \cite{ori}, it makes the training more intricate while not resulting in substantial improvement.

We train using a loss that also incorporates the error in the spatial gradients, similarly to \cite{Aittala}
\begin{equation}
    \textrm{GradLoss}(\mathbf{Z}, \hat{\mathbf{Z}}) = \frac{1}{10}||\mathbf{Z} - \hat{\mathbf{Z}}||_2^2 + ||\nabla \mathbf{Z} - \nabla \hat{\mathbf{Z}}||_2^2.
    \label{eq:grad_loss}
\end{equation}
Here $\hat{\mathbf{Z}}$ and $\mathbf{Z}$ are the network's prediction and the target, respectively, and $\nabla$ denotes the horizontal and vertical finite differences. The second term encourages the edges of the output image to be  similar to those of the target image.

\section{Experimental Study}
\label{sec:results}
\noindent \textbf{Data:} Since to our best knowledge no publicly available dataset targets the derverberation task based on ad-hoc microphone arrays, we created such a dataset to train and evaluate our network. The dataset comprises four scenarios: (1) all microphones are far from the speaker (2) all microphones are near the speaker (3) all microphones are randomly placed (4) $M-1$ microphones are far from the speaker and another one is close by. We refer to the latter case as the \emph{Winning Ticket} scenario, since a successful technique will result in an enhanced signal which is at least as good as the near-microphone signal. Two versions of the dataset were generated; a noiseless variant which we call BIUREV, and a noisy counterpart named BIUREV-N.

For each recording, the room's length and width were drawn such that the smaller dimension was in the range $\mathcal{U}(4, 7)$ metres, where $\mathcal{U}$ is a uniform distribution. Subsequently, an aspect ratio was drawn from $\mathcal{U}(1, 1.5)$. The height of the room had a constant value of 2.7m. The reverberation time of the room ($T_{60}$) was also randomly chosen as one of the values $\{0.2, 0.4, 0.7, 1\}$ seconds. The sound source was placed at a height of 1.75m and the microphones at 1.6m. The sound source and the microphones were at least 0.5m away from the walls.

The threshold distance that distinguishes the ``near" and ``far" conditions is called the \emph{critical distance}, denoted here by $d_{\textrm{crit}}$. It is determined by the room's volume and $T_{60}$. The speaker-near microphone and speaker-far microphone distances were drawn from $\mathcal{U}(0.2, d_{\textrm{crit}})$ and  $\mathcal{U}(2d_{\textrm{crit}}, 3)$, respectively. For the third scenario, i.e. all-random placement, the distance for all microphones was drawn from $\mathcal{U}(0.2, 3)$. For BIUREV-N, noise signals at \ac{SNR} of 20dB were independently added to each reveberant microphone signal. It was generated by filtering white Gaussian noise with an auto-regressive filter with order 1 to emphasize the lower frequency band. 

For training, 7861 reverberant speech recordings were generated with random microphone positioning. Validation data comprised 742 utterances for Near and Far conditions, and 1088 test utterances were generated for each scenario. The rooms and positions varied from training to test. Clean recordings, sampled at 16KHz, were taken from the REVERB Challenge \cite{REVERB}, and the \acp{RIR} were generated using the gpuRIR package \cite{gpuRIR}.

For calculating the \ac{STFT}, a Hanning window with $K=512$ and 75\% overlap between successive frames were used for analysis and synthesis. During training, slices of size $256 \times F$ were randomly sampled from the reverberant log-spectrum across the $M$ microphones together with the corresponding slice from the clean spectrogram.  In order to confine the clean and noisy spectrograms to $[-1,1]$, the minimum and maximum values over the entire training dataset were calculated before the training stage. These values were also used to map the network's output back to the legitimate range of values of a clean spectrogram.

\noindent \textbf{Experimental setup: }
We train and test our network on BIUREV and BIUREV-N. For the more challenging noisy dataset, it is trained on eight and four microphones together by randomly drawing the number of microphones to be used for each mini-batch. Two additional \ac{DSS} networks were trained on eight and four microphones separately. For the BIUREV dataset, only an eight microphones network was trained.

\begin{table*}[t]
    
    \centering
    \caption{Results for noiseless reverberant speech for the different scenarios}
    \label{table:BIUREV}
    \vspace{-3pt}
    \begin{subtable}[h]{0.45\textwidth}
        \centering
        \setlength{\belowcaptionskip}{.1cm}
        \caption{Far}
        \label{table:far}
        \resizebox{0.9\linewidth}{!}{%
        \begin{tabular}{@{}lcccc@{}}
        \toprule
         & Scene-Agnostic & \multicolumn{1}{c}{CD$\downarrow$} & \multicolumn{1}{c}{FWSegSNR$\uparrow$} & \multicolumn{1}{c}{PESQ$\uparrow$} \\
         \cmidrule(lr){3-3} \cmidrule(lr){4-4} \cmidrule(lr){5-5}
        {reverberant} & - & 4.72 & 4.54 & 1.37 \\
        \hline
        {8 mics.} & \xmark & 2.45 & 12.10 & 2.15 \\
        {8 mics.} & \cmark & \textbf{2.44} & \textbf{12.15} & 2.07 \\
        {WPE \cite{WPE}} & \cmark & 3.36 & 8.32 & \textbf{2.30} \\
         \bottomrule
        \end{tabular}}%
    \end{subtable}%
    \begin{subtable}[h]{0.45\textwidth}
        \centering
        \setlength{\belowcaptionskip}{.1cm}
        \caption{Near}
        \label{table:near}
        \resizebox{0.9\linewidth}{!}{%
        \begin{tabular}{@{}lcccc@{}}
        \toprule
         & Scene-Agnostic & \multicolumn{1}{c}{CD$\downarrow$} & \multicolumn{1}{c}{FWSegSNR$\uparrow$} & \multicolumn{1}{c}{PESQ$\uparrow$} \\
         \cmidrule(lr){3-3} \cmidrule(lr){4-4} \cmidrule(lr){5-5}
        {reverberant} & - & 3.35 & 10.40 & 1.84 \\
        \hline
        {8 mics.} & \xmark & 1.95 & 15.04 & 2.88 \\
        {8 mics.} & \cmark & \textbf{1.76} & 15.86 & 3.07 \\
        {WPE \cite{WPE}} & \cmark & 1.86 & \textbf{16.22} & \textbf{3.22} \\
         \bottomrule
        \end{tabular}}%
    \end{subtable}%

    \begin{subtable}[h]{0.45\textwidth}
        \centering
        \setlength{\belowcaptionskip}{.1cm}
        \caption{Random}
        \label{table:random}
        \resizebox{0.9\linewidth}{!}{%
        \begin{tabular}{@{}lcccc@{}}
        \toprule
         & Scene-Agnostic & \multicolumn{1}{c}{CD$\downarrow$} & \multicolumn{1}{c}{FWSegSNR$\uparrow$} & \multicolumn{1}{c}{PESQ$\uparrow$} \\
         \cmidrule(lr){3-3} \cmidrule(lr){4-4} \cmidrule(lr){5-5}
        {reverberant} & - & 4.4 & 5.65 & 1.48 \\
        \hline
        {8 mics.} & \xmark & 2.04 & 13.96 & 2.73 \\
        {8 mics.} & \cmark & \textbf{2.01} & \textbf{14.28} & \textbf{2.81} \\
        {WPE \cite{WPE}} & \cmark & 2.95 & 10.15 & 2.54 \\
         \bottomrule
        \end{tabular}}%
    \end{subtable}%
    \begin{subtable}[h]{0.45\textwidth}
        \centering
        \setlength{\belowcaptionskip}{.1cm}
        \caption{Winning Ticket}
        \label{table:winning}
        \resizebox{0.9\linewidth}{!}{%
        \begin{tabular}{@{}lcccc@{}}
        \toprule
         & Scene-Agnostic & \multicolumn{1}{c}{CD$\downarrow$} & \multicolumn{1}{c}{FWSegSNR$\uparrow$} & \multicolumn{1}{c}{PESQ$\uparrow$} \\
         \cmidrule(lr){3-3} \cmidrule(lr){4-4} \cmidrule(lr){5-5}
        {reverberant} & - & 3.36 & 10.44 & 1.81 \\
        \hline
        {8 mics.} & \xmark & 2.02 & 13.94 & 2.80 \\
        {8 mics.} & \cmark & 1.91 & 14.67 & 2.93 \\
        {WPE \cite{WPE}} & \cmark & \textbf{1.84} & \textbf{16.32} & \textbf{3.22} \\
         \bottomrule
        \end{tabular}}%
    \end{subtable}%
\end{table*}

\noindent \textbf{Baselines and evaluation criteria: }
The criteria used for comparison are \ac{CD}, \ac{PESQ} and \ac{FWSegSNR}. 

For comparison on BIUREV-N, the two baselines are the powerful single microphone paradigm \cite{ori} and its scene-aware microphone array extension that concatenates the spectrograms in the feature dimension. We used our own implementation of the network described in \cite{ori}, and trained it with GradLoss (\ref{eq:grad_loss}). Note that since each \ac{DSS} layer uses two convolution layers, our network has twice as many parameters compared to \cite{ori}. For fairness, we made sure both networks used the same number of parameters. Benchmarks for the corrupted speech were calculated on the closest microphone. Single microphone enhancement was also applied to the nearest microphone, except for the Winning Ticket scenario. 

For BIUREV, the scene-specific network based on \cite{ori} was used, too. We also compared the proposed architecture to another scene-agnostic technique, i.e. WPE \cite{WPE}, which is not based on prior training. It did not feature in the BIUREV-N performance test, since its result had a significant level of noise at the output.

\noindent \textbf{Discussion:}
Table \ref{table:BIUREV-N} provides the quantitative results for BIUREV-N. The scene-aware and scene-agnostic networks were tested on four and eight microphones. As can be seen, for all scenarios and networks the performance consistently improves with the number of microphones across all benchmarks. Another important observation is that for most cases in the scene-agnostic paradigm, the combined model trained on both eight and four microphones is on par or even slightly better compared to the respective models trained separately. In the Winning Ticket scenario, substantial improvement is achieved with respect to the Far case.

The scene-agnostic networks are superior to the scene-aware ones across all conditions, except for the Far case where they perform equally well. Remarkably, in some cases using four microphones with the scene-agnostic network leads to better scores than with eight microphones under the scene-aware setup. We conjecture that the scene-aware setup still yields competitive results since the phase data, which is closely related to the speaker-microphone placement, is not incorporated in the spectrograms. 

Benchmarks for the BIUREV dataset are presented in Table \ref{table:BIUREV}. The scene-agnostic and scene-aware networks showcase the same trend observed for the noisy dataset in Table \ref{table:BIUREV-N}. For the Far condition the scene-aware network results in equivalent performance with better PESQ score.  For all other cases, the proposed architecture emerges advantageous. 

When compared to WPE, our network is favourable for Random and Far scenarios. In the latter, WPE exhibits better PESQ, but due to its relatively low FWSegSNR, the enhanced signal is still reverberated. For Near condition, WPE obtains the best PESQ and FWSegSNR whereas the scene-agnostic network leads in the CD criterion. In the case of Winning Ticket, WPE achieves the best scores due to its mechanism. In WPE, the reverberation in one microphone is estimated based on all microphones. This means that for an eight-microphone array, WPE outputs eight enhanced signals, and the best one is picked. In the Winning Ticket case, the best signal is obtained from the near microphone signal.

Ultimately, the scene-agnostic architecture seems to be the most successful and practical with respect to the baselines, for two reasons. Conversely to WPE which is considerably susceptible to noise, it can cope with noisy and  noiseless cases. In addition, it can accommodate different number of microphones in a single network unlike the scene-aware baseline. Moreover, it results in better enhancement quality. The interested reader is encouraged to listen to audio samples of dereverberated recordings using the different methods.\footnote{\texttt{www.eng.biu.ac.il/gannot/speech-enhancement}}


\section{Conclusion}
We presented an \ac{NN} architecture which is oblivious to the number and positions of microphones in a microphone array. The suggested architecture builds upon the Deep Sets framework and \ac{DSS} layers. In contrast, the common approach that concatenates the multi-microphone signals across the channel axis assumes prior knowledge on the constellation and therefore hinders its applicability to the scene-agnostic setup. The new architecture was used to enhance reverberant log-spectrum from multiple noisy observations, and tested on novel datasets that carefully examine different facets of our paradigm. The objective and subjective tests showed that even though the baselines are extremely successful, the proposed method improves the performance even with fewer microphones. Moreover, our method lends itself to conveniently accommodate different number of microphones in a single network, and generalises well for unseen combinations of microphones.  


 \vfill\pagebreak
\balance
\bibliographystyle{IEEEbib}
\bibliography{refs}

\end{document}